\documentclass[notitlepage,aps,prl,reprint,superscriptaddress]{revtex4-1}

\renewcommand{\thesection}{\Roman{section}}

\renewcommand{\figurename}{FIG.}

\usepackage{qcircuit}
\usepackage{mathtools}
\usepackage[english]{babel}
\usepackage[utf8]{inputenc}
\usepackage{amsmath}
\usepackage{amsfonts}
\usepackage{amssymb}
\usepackage{graphicx}
\usepackage{bbold}
\usepackage{epstopdf}
\usepackage{hyperref}
\renewcommand*{\Re}{\operatorname{Re}} 
 
\DeclarePairedDelimiter\bra{\langle}{\rvert}
\DeclarePairedDelimiter\ket{\lvert}{\rangle}

\newcommand{\x}{\hat{x}}
\newcommand{\appropto}{\mathrel{\vcenter{
  \offinterlineskip\halign{\hfil$##$\cr
    \propto\cr\noalign{\kern2pt}\sim\cr\noalign{\kern-2pt}}}}}

\begin{document}
\renewcommand{\figurename}{FIG.}
\renewcommand{\tablename}{TABLE}
\title{Cubic phase gates are not suitable for non-Clifford operations on GKP states
}
\author{Jacob Hastrup}
\affiliation{Center for Macroscopic Quantum States (bigQ), Department of Physics, Technical University of Denmark, Building 307, Fysikvej, 2800 Kgs. Lyngby, Denmark}
\email{jhast@fysik.dtu.dk}
\author{Mikkel V. Larsen}
\affiliation{Center for Macroscopic Quantum States (bigQ), Department of Physics, Technical University of Denmark, Building 307, Fysikvej, 2800 Kgs. Lyngby, Denmark}
\author{Jonas S. Neergaard-Nielsen}
\affiliation{Center for Macroscopic Quantum States (bigQ), Department of Physics, Technical University of Denmark, Building 307, Fysikvej, 2800 Kgs. Lyngby, Denmark}
\author{Nicolas C. Menicucci}
\affiliation{Centre for Quantum Computation \& Communication Technology, School of Science, RMIT University, Melbourne, VIC 3000, Australia}
\author{Ulrik L. Andersen}
\affiliation{Center for Macroscopic Quantum States (bigQ), Department of Physics, Technical University of Denmark, Building 307, Fysikvej, 2800 Kgs. Lyngby, Denmark}


\begin{abstract}
With the Gottesman-Kitaev-Preskill (GKP) encoding, Clifford gates and error correction can be carried out using simple Gaussian operations. Still, non-Clifford gates, required for universality, require non-Gaussian elements. In their original proposal, GKP suggested a particularly simple method of using a single application of the cubic phase gate to perform the logical non-Clifford T-gate. Here we show that this cubic phase gate approach performs extraordinarily poorly, even for arbitrarily large amounts of squeezing in the GKP state. Thus, contrary to common belief, the cubic phase gate is not suitable for achieving universal fault-tolerant quantum computation with GKP states. 
\end{abstract}
\date{\today}

\maketitle

\textit{Introduction}.---The Gottesman-Kitaev-Preskill (GKP) encoding of a qubit into a quantum harmonic oscillator \cite{gottesman2001encoding} is a particularly promising approach towards fault-tolerant quantum computation. In particular, all Clifford operations, including error correction, can be implemented using only Gaussian operations along with a supply of ancillary GKP basis states. Furthermore, the GKP encoding scheme has been shown to outperform other bosonic codes against loss \cite{albert2018performance,noh2018quantum}, which is the dominant noise source in most bosonic systems. For these reasons, the GKP encoding has gained much interest in recent years, both theoretically and experimentally, across multiple experimental platforms. Most prominently, the states have been generated in trapped-ion \cite{fluhmann2019encoding} and microwave cavity platforms \cite{campagne2019quantum}. Furthermore, in the optical regime, large 2-dimensional cluster states have been produced \cite{larsen2019deterministic,asavanant2019generation}, which enable scalable fault-tolerant measurement-based quantum computation when combined with high quality GKP states \cite{menicucci2014fault}.

A critical step towards universality is the ability to perform non-Clifford operations on the encoded qubits. For GKP qubits, two different approaches for non-Clifford operations were proposed in the original paper \cite{gottesman2001encoding}. The first approach is to use logical magic states, such as the encoded Hadamard eigenstate, to implement the non-Clifford T-gate via gate teleportation. Such magic states can be distilled using only the computational basis states and Gaussian operations \cite{baragiola2019all}, or they can be generated directly using non-Gaussian resources such as photon counting \cite{gottesman2001encoding,menicucci2014fault,su2019conversion} or coupling to a two-level system \cite{campagne2019quantum,hastrup2019measurement}. The second approach is to apply a single cubic phase gate in combination with Gaussian operations \cite{gottesman2001encoding}. In principle, the cubic phase gate enables universal control of the oscillator \cite{lloyd1999quantum}, including any desired operations on the GKP state. However,
a significant overhead, requiring many applications of the cubic phase gate, is typically required to approximate most non-Gaussian operations well with cubic phase gates. The promise that a single application would suffice to implement a logical non-Clifford operation therefore strongly motivates the development of cubic phase gates for applications with GKP states. GKP also showed that the cubic phase gate could be implemented using a cubic phase state and teleportation with Gaussian operations. Since the ideal cubic phase state is nonphysical, requiring infinite energy, this teleportation-based technique is always approximate. Still, even when using an ideal cubic phase gate the approach is only suitable for GKP states with an asymmetric noise distribution, as was pointed out by GKP in their original paper \cite{gottesman2001encoding}.

In this paper we analyse the details of the cubic phase gate approach and show explicitly that it performs surprisingly poorly, unless the GKP state is prepared with an unrealistic noise distribution. We consider only a perfect implementation of the cubic phase gate in order to discount any imperfections e.g. from finite energy cubic phase states. The results presented here thus represent a best-case scenario for the cubic phase gate approach. The poor performance is therefore solely due to the intrinsic and unavoidable noise present in the GKP states. We also compare the performance to that achieved using a GKP-encoded magic state via gate teleportation, demonstrating that the magic state offers a significantly better approach. 

\textit{Preliminaries}.---We consider a bosonic mode of a quantum harmonic oscillator with position and momentum quadrature operators $\hat{x}$ and $\hat{p}$ satisfying $[\hat{x},\hat{p}]=i$ with vacuum variance $\text{Var}(\hat{x})=\text{Var}(\hat{p})=1/2$. A detailed review of GKP states and their error-correcting properties can be found elsewhere \cite{gottesman2001encoding,tzitrin2020progress,terhal2020towards}. We focus on the approximate square GKP states consisting of a sum of equispaced position-squeezed states under a Gaussian envelope:
\begin{align}
    \ket{0_L} &\propto \sum_{s\in \mathbb{Z}}e^{-\pi (2s)^2\Delta_p^2/2}\int dx\, e^{-\frac{(x-2s\sqrt{\pi})^2}{2\Delta_x^2}}\ket{x} \\
    \ket{1_L} &\propto \sum_{s\in \mathbb{Z}}e^{-\pi (2s+1)^2\Delta_p^2/2}\int dx\, e^{-\frac{(x-(2s+1)\sqrt{\pi})^2}{2\Delta_x^2}}\ket{x},
\end{align}
where ``$L$'' denotes logical qubit states, $\Delta_x$ and $\Delta_p$ quantifies the amount of squeezing, or noise, in the $x$- and $p$-quadratures respectively, and $\ket{x}$ are the position eigenstates, i.e. $\hat{x}\ket{x}=x\ket{x}$. Note that the squeezing is that of each of the peaks of the state, not the overall state, i.e. each peak in the $x$- and $p$-quadrature has a measured variance of $\Delta_x/2$ and $\Delta_p/2$ respectively. Hence, $\Delta_x$ and $\Delta_p$ can be arbitrarily low simultaneously \cite{duivenvoorden2017single}. Numerical values of the squeezing are often expressed in decibels (dB) as $-10\log_{10}(\Delta^2)$. Importantly, a small amount of noise in $p$ enforces a wide envelope in $x$ and vice versa. In the limit of infinite squeezing, i.e. $(\Delta_x,\Delta_p)\rightarrow (0,0)$, the position wave functions of the computational basis states approach Dirac combs with spacing $2\sqrt{\pi}$. 

\begin{figure}
    \centering
    \includegraphics{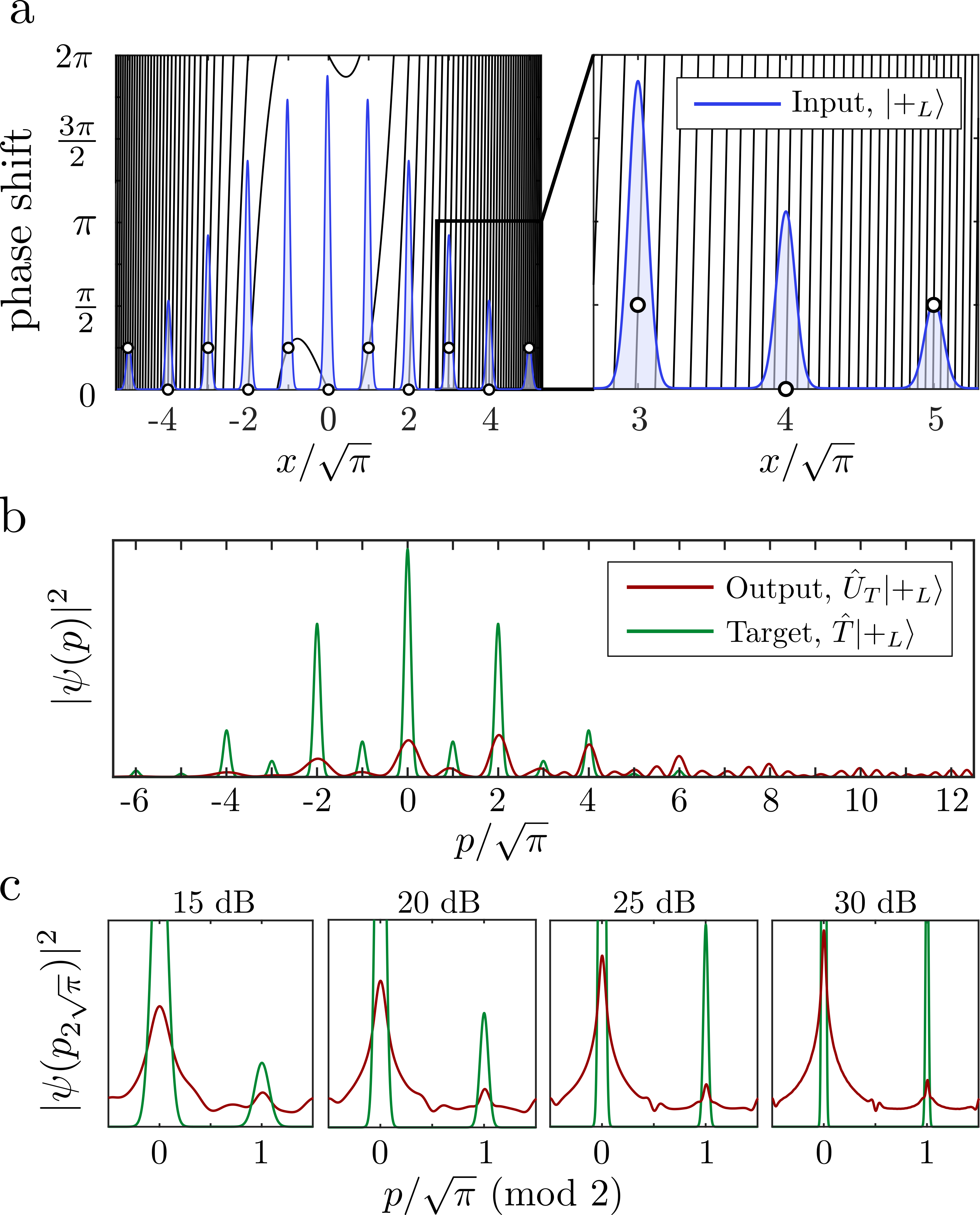}
    \caption{(a): Position ($x$) probability density (blue) of $\ket{+_L}=(\ket{0_L}+\ket{1_L})/\sqrt{2}$ with $\Delta_x=\Delta_p=15$ dB. Also plotted in black is the phase shift imposed by the operator $\hat{U}_T$ (Eq. \eqref{eq:UT}). The white circles show the value of the phase shift at integer multiples of $\sqrt{\pi}$. The right window is a zoom in on the peaks further from the origin. These peaks experience a large phase variation across their width, greatly limiting the performance of $\hat{U}_T$ as a T-gate. (b): Momentum ($p$) probability density of $\hat{U}_T\ket{+_L}$ (red) compared to the target state $\hat{T}\ket{+_L}=(\ket{0_L} + e^{i\pi/4}\ket{1_L})/\sqrt{2}$ (green). (c): Momentum probability densities summed over values of $p$ mod $2\sqrt{\pi}$, i.e. $|\psi(p_{2\sqrt{\pi}})|^2 = \sum_s|\psi(p + 2s\sqrt{\pi})|^2$ for 15-30 dB squeezing, for $\hat{U}_T\ket{+_L}$ (red) and $\hat{T}\ket{+_L}$ (green). The ratio of the heights of the green peaks remain constant as the squeezing is increased.}
    \label{fig:Fig1}
\end{figure}

We now consider how to implement the non-Clifford T-gate, also known as the $\pi/8$-gate,
\begin{equation}
    \hat{T}=\ket{0_L}\bra{0_L} + e^{i\pi/4}\ket{1_L}\bra{1_L},
\end{equation}
which, when combined with the Clifford gate set, constitutes a universal gate set. GKP proposed to use a cubic phase gate in combination with shearing and displacement to implement the T-gate:
\begin{equation}
    \hat{U}_T=\exp\Bigg[i\frac{\pi}{4}\bigg\{ 2 \underbrace{\left(\frac{\x}{\sqrt{\pi}}\right)^3}_{\textrm{cubic phase}}+\underbrace{\left(\frac{\x}{\sqrt{\pi}}\right)^2}_{\textrm{shear}}-\!\!\!\!\underbrace{2\frac{\x}{\sqrt{\pi}}}_{\;\;\textrm{displacement}}\!\!\!\!\!\!\!\!\bigg\}\Bigg]. \label{eq:UT}
\end{equation}
One can check that this gate applies a $\pi/4$ phase shift to peaks positioned at odd multiples of $\sqrt{\pi}$ since 
\begin{equation}
    2x^3+x^2-2x \equiv \begin{cases} 0 \;\textrm{(mod $8$)} &\text{for even $x$,} \\
    1 \;\textrm{(mod $8$)} &\text{for odd $x$.}\end{cases}
\end{equation}
For ideal GKP states with support only at integer multiples of $\sqrt{\pi}$, $\hat{U}_T$ thus acts as a perfect T-gate. However, approximate GKP states also have support outside these grid points where $\hat{U}_T$ does not exactly apply a 0 or $\pi/4$ phase shift. This is illustrated in Fig.\ \ref{fig:Fig1}a, showing the GKP wave function of $\ket{+_L} = (\ket{0_L} + \ket{1_L})/\sqrt{2}$, along with the polynomial in $\hat{U}_T$ modulo $2\pi$. Only for the peaks close to $x=0$ is the phase shift approximately correct over the width of each peak. For peaks further from the origin, e.g. for $|x|/\sqrt{\pi}>3$ in the case of Fig.\ \ref{fig:Fig1}a, there is a large phase variation across the peaks. The peaks far from the origin thus effectively experience a random phase shift. This results in an increased amount of noise in the $p$-quadrature, which is seen in Fig.\ \ref{fig:Fig1}b. Furthermore, the output state is highly asymmetric with a long tail at positive $p$. This is because $\hat{U}_T$ can be interpreted as a momentum displacement with an $x$-dependent displacement magnitude scaling as $x^2$. Peaks at large $|x|$ thus get displaced to large $p$-values.

How does this noise behave as we increase the amount of squeezing in the state? On one hand, as we decrease $\Delta_x$ the width of each peak decreases, thus decreasing the total phase variation across each peak. On the other hand, as we decrease $\Delta_p$, peaks further from the origin appear due to the Fourier relation between $\hat{x}$ and $\hat{p}$. These new peaks now experience a larger phase variation as seen in Fig.\ \ref{fig:Fig1}a. It turns out that for $\Delta_x=\Delta_p$, new peaks appear at a rate comparable to the rate at which they narrow, such that the gate fidelity does \textit{not} converge to 1 when increasing the squeezing. This is qualitatively illustrated in Fig.\ \ref{fig:Fig1}c, showing the momentum probability density summed over values of $p$ modulo $2\sqrt{\pi}$. Even as the squeezing approaches very large values, the probability density retains a non-zero noise floor with significant support outside integer multiples of $\sqrt{\pi}$. 

This poor noise distribution was pointed out in the original paper by GKP \cite{gottesman2001encoding}, stating that one needs to ensure that $\Delta_x\ll\Delta_p$ in order to use $\hat{U}_T$ as a logical T-gate. However, this condition is highly impractical to maintain. For example, the logical Hadamard gate, which is implemented by a $\pi/2$ phase rotation,  $\hat{U}_H=\exp(i\pi/4(\hat{x}^2 + \hat{p}^2))$, also swaps the noise of the $x$- and $p$-quadratures, i.e. swapping $\Delta_x\Leftrightarrow \Delta_p$. Additionally, since we want low noise in both quadratures, i.e. both $\Delta_x \ll 1$ and $\Delta_p \ll 1$, the condition $\Delta_x\ll\Delta_p$ requires an extremely low value of $\Delta_x$, which will be difficult, if not impossible, to produce and maintain experimentally. 
In the following we quantify the above considerations numerically and analytically. 

\begin{figure}
    \centering
    \includegraphics{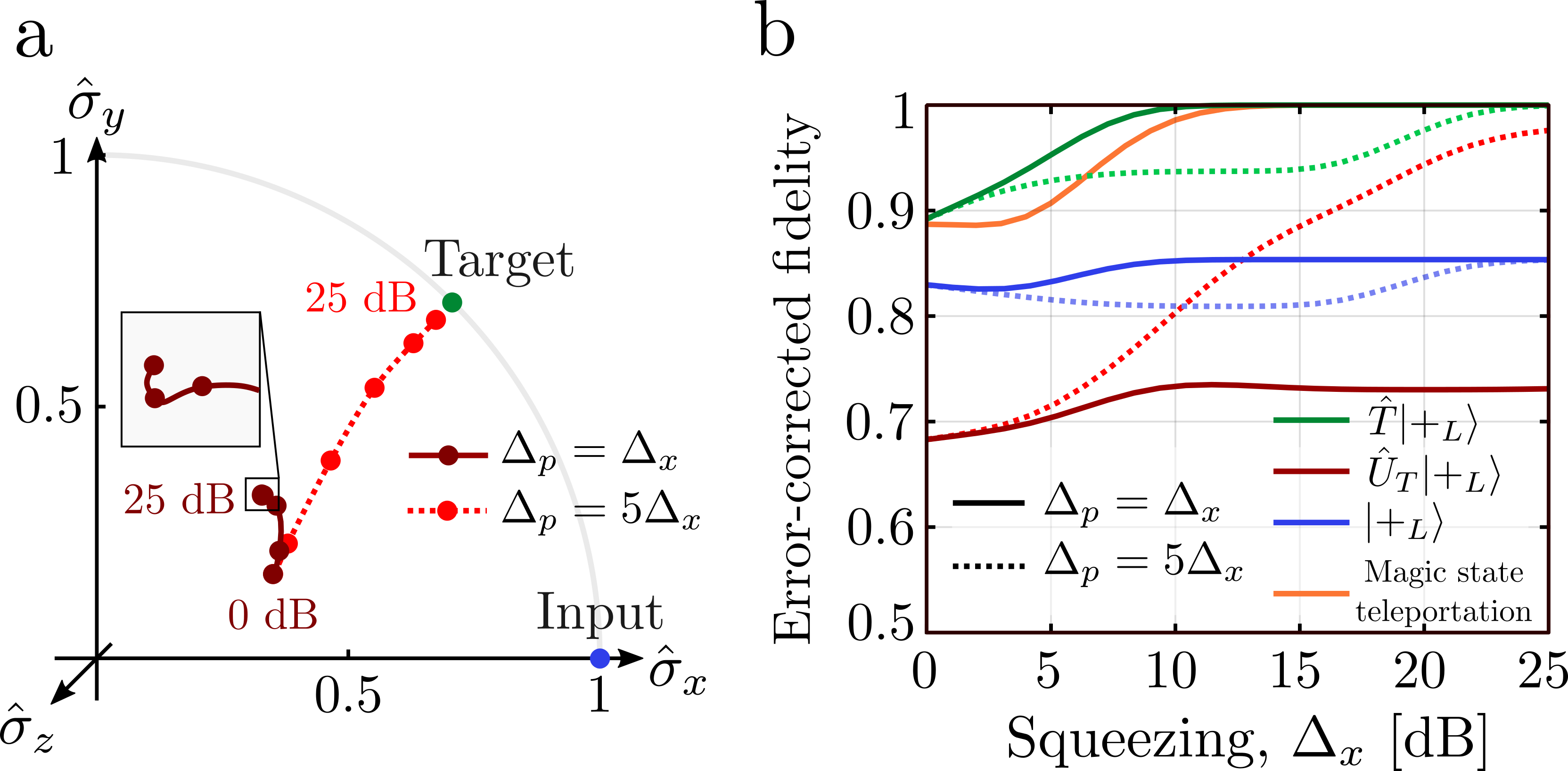}
    \caption{(a): $x$-$y$ plane of the Bloch sphere with the average qubit output of the GKP error correction scheme with $\hat{U}_T\ket{+_L}$ as input. Each dot corresponds to a step of 5 dB in the squeezing level of the $x$-quadrature. (b): Fidelity, $F=\bra{T}\rho\ket{T}$ with the target qubit state, $\ket{T}=(\ket{0}+e^{i\pi/4}\ket{1})/\sqrt{2}$, for the qubit with density matrix, $\rho$, of the GKP error corrected qubits of $\hat{U}_T\ket{+_L}$ (red), $\hat{T}\ket{+_L}$ (green) and $\ket{+_L}$ (blue), calculated for both $\Delta_p=\Delta_x$ (solid) and $\Delta_p=5\Delta_x$ (dotted) in the initial $\ket{+_L}$ state, as well as the fidelity obtained by the magic state approach with gate teleportation \cite{gottesman2001encoding} (orange).}
    \label{fig:Fig2}
\end{figure}

\textit{Error-corrected fidelity}.---One could try to perform GKP error correction to correct the noise generated by $\hat{U}_T$. Here we consider the best-case scenario of perfect error correction i.e. using ideal GKP ancillas with infinite squeezing in order to discount any imperfections in the error correction protocol. Such perfect error correction corresponds to a measurement-dependent displacement followed by a projection of the state into the ideal 2-dimensional GKP subspace \cite{baragiola2019all}. The output state is thus described by a qubit which depends on the input state and the syndrome measurement outcome. Averaging the output states over all syndrome outcomes, we generally obtain a mixed state described by a qubit density matrix $\rho$. Fig.\ \ref{fig:Fig2}a shows the position of $\hat{U}_T\ket{+_L}$ on the Bloch sphere after error correction for various squeezing levels. For decreasing $\Delta_x=\Delta_p$ the state converges to a point well-inside the Bloch sphere, and not to the target state $\ket{T}=(\ket{0} + e^{i\pi/4}\ket{1})/\sqrt{2}$. The situation is improved by considering an asymmetric noise distribution, e.g. when $\Delta_p=5\Delta_x$ (corresponding to 14dB less squeezing in the $p$-quadrature), as expected. Fig.\ \ref{fig:Fig2}b shows the fidelity, $F=\bra{T}\rho\ket{T}$, to the target state. For $\Delta_x=\Delta_p$ we find that the fidelity to the target state is lower for $\hat{U}_T\ket{+_L}$ (red line) compared to $\ket{+_L}$ (blue line) for all squeezing levels. Thus for symmetric noise, $\hat{U}_T$ is a worse T-gate than the identity gate! Again the situation is improved by asymmetric noise, but the fidelity increases slowly with $\Delta_x$ compared to an ideal T-gate (green), and the fidelity still does not converge to 1. For comparison we also plot the fidelity obtained when using GKP-encoded magic states to teleport the T-gate, as proposed in \cite{gottesman2001encoding} (orange). The squeezing level of the magic state ancilla is equal to that of the input states. The finite squeezing of the ancilla results in a performance which is slightly lower than the ideal T-gate (green line), but the approach is significantly better than the cubic phase gate approach.

\begin{figure}
    \centering
    \includegraphics{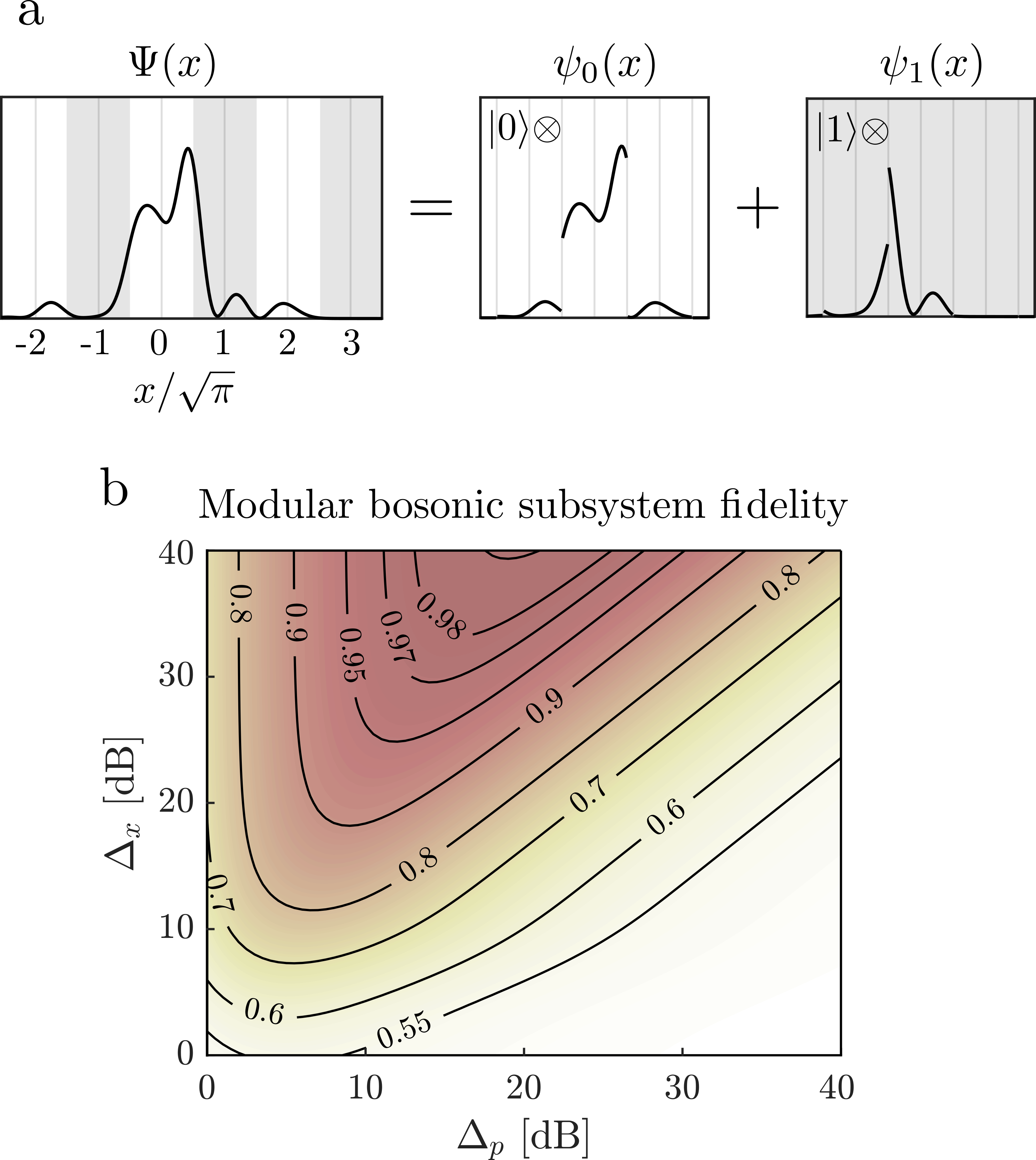}
    \caption{(a): Illustration of the modular bosonic subsystem decomposition technique \cite{pantaleoni2020modular} for a random wave function, $\Psi(x)$. The position wave function is decomposed by binning it around even and odd multiples of $\sqrt{\pi}$ and stitching the bins together to form two new wave functions, $\psi_0$ and $\psi_1$. (b): Fidelity with the target qubit state, $(\ket{0}+e^{i\pi/4}\ket{1})/\sqrt{2}$ of the qubit part of the input state $\hat{U}_T\ket{+_L}$, after tracing out the continuous parts with the modular bosonic subsystem decomposition technique
    for different values of $\Delta_x$ and $\Delta_p$.}
    \label{fig:Fig3}
\end{figure}

\textit{Modular bosonic subsystem fidelity}.---An alternative framework for reducing a bosonic state to a 2-dimensional GKP qubit was recently proposed by Pantaleoni et al. \cite{pantaleoni2020modular}. The idea is to decompose an arbitrary bosonic state, $\ket{\Psi}$, into a qubit part and a continuous part, i.e.
\begin{equation}
    \ket{\Psi} = \ket{0}\otimes\ket{\psi_0} + \ket{1}\otimes\ket{\psi_1}.
\end{equation}
The decomposition is done by binning the wave function around even and odd multiples of $\sqrt{\pi}$ and stitching the bins together to form two new wave functions $\ket{\psi_0}$ and $\ket{\psi_1}$, as illustrated in Fig.\ \ref{fig:Fig3}a. Tracing out the continuous part we are left with a qubit state which contains the logical information of the state. Further details on this technique can be found in the supplementary material and in Ref.\ \cite{pantaleoni2020modular}. Using this method we can again analyse the fidelity of $\hat{U}_T\ket{+_L}$ with the target state in the qubit subspace, thus providing a complementary figure of merit. The result is shown in Fig.\ \ref{fig:Fig3}b. Again we observe that in order to obtain a high fidelity, we need an excess amount of squeezing in the $x$-quadrature. For example, to achieve a fidelity of 0.95, we require $\Delta_x>25$ dB, which is significantly more than the squeezing thresholds set for fault-tolerance using magic states to implement a T-gate \cite{menicucci2014fault,fukui2018high,Yamasaki2020polylog,noh2020fault}. For $\Delta_x=\Delta_p$ we again observe a convergence in the fidelity below 1. In fact, in the limit of infinite squeezing, $(\Delta_x,\Delta_p)\rightarrow(0,0)$, one can derive the following analytical result (see supplementary information):
\begin{equation}
    F = \frac{1}{2} + \frac{1}{2}\frac{1}{\sqrt{1+\left(\frac{3\Delta_x}{2\Delta_p}\right)^2}}. \label{eq:Fidelity}
\end{equation}
Thus the fidelity is bounded, confirming the poor performance even in the limit of infinite squeezing. In particular, for the realistic case of $\Delta_x = \Delta_p$ we get $F = 1/2 + 1/\sqrt{13}\approx 0.78 < 1$. To obtain a higher fidelity we require $\Delta_x\ll\Delta_p$ in which case Eq.\ \eqref{eq:Fidelity} reduces to
\begin{equation}
    F \approx 1 - \left(\frac{3}{4}\frac{\Delta_x}{\Delta_p}\right)^2, \qquad \text{for $\Delta_x\ll \Delta_p$},
\end{equation}
which goes to 1 in the limit $\Delta_x/\Delta_p\rightarrow0$, as expected. Again, such unbalanced noise ratio is not realistic to maintain during a calculation since the logical Hadamard gate (i.e. a Fourier transformation) swaps the noise between the quadratures. 

\textit{Conclusion}.---We have analysed the performance of the GKP T-gate implemented via the cubic phase gate. All calculations were performed assuming a perfect cubic phase gate, and thus represents a best-case scenario compared to approximate implementations using cubic phase states. In the case of square GKP states with equal squeezing in the position and momentum quadrature we have explicitly shown that the noise generated by the T-gate is detrimental, \textit{for all squeezing levels}. The poor performance is solely due to the noise inherent in all physical GKP states, and cannot be circumvented through error correction. It can in principle be mitigated by using highly asymmetrical GKP states, but this strategy is not compatible with other manipulations of the state, such as the logical Hadamard gate (i.e. a Fourier transform), which swaps the noise of the position and momentum quadratures. Although we have focused our analysis on square GKP states, it is clear that other GKP grids, e.g. hexagonal GKP states, will suffer from the same issues. That is, the rapid variation of the phase applied by the cubic phase gate with respect to $x$ will eventually cause detrimental issues at large $|x|$ for any grid. In fact, one can generalise this to any operator consisting of a finite polynomial in $\hat{x}$, as the derivative of any finite polynomial is unbounded. Instead, the T-gate should be implemented using GKP-encoded magic states, and efforts towards optical GKP-based quantum computation should be focused on the generation of GKP states and not on the development of cubic phase gates or cubic phase states.
\\

This project was supported by the Danish National Research Foundation through the Center of Excellence for Macroscopic Quantum States (bigQ, DNRF0142) and by the Australian Research Council Centre of Excellence for Quantum Computation and Communication Technology (Project No. CE170100012).

\section*{References}
\bibliography{References}

\clearpage
 \setcounter{section}{0}
    \renewcommand{\thesection}{S\arabic{section}}
    \setcounter{equation}{0}
    \renewcommand{\theequation}{S\arabic{equation}}
    \setcounter{figure}{0}
    \renewcommand{\thefigure}{S\arabic{figure}}
    
\begin{widetext}
\section*{Supplementary Material}

Here we derive Eq. \eqref{eq:Fidelity} from the main text. As explained in the main text, the main idea behind the modular bosonic subsystem decomposition technique \cite{pantaleoni2020modular} is to divide the continuous Hilbert space of the bosonic mode into a 2-dimensional part, describing the logical content of the GKP qubit, and a continuous part. Tracing out the continuous part leaves us with a mixed qubit state, which can be analysed e.g. in terms of fidelity to a target qubit. The decomposition is done by binning the $x$-quadrature wave function into bins around even and odd multiples of $\sqrt{\pi}$, as illustrated in Fig.\ \ref{fig:Fig3}a. An arbitrary pure state can thus be written as:
\begin{equation}
    \ket{\Psi} = \ket{0}\otimes\ket{\psi_0} + \ket{1}\otimes\ket{\psi_1}
\end{equation}
where 
\begin{subequations}
\begin{align}
    \ket{\psi_0} &= \sum_{s\in\mathbb{Z}}\int_{(2s-1/2)\sqrt{\pi}}^{(2s+1/2)\sqrt{\pi}}dx\,\Psi(x)\ket{x}\\
    \ket{\psi_1} &= \sum_{s\in\mathbb{Z}}\int_{(2s-1/2)\sqrt{\pi}}^{(2s+1/2)\sqrt{\pi}}dx\,\Psi(x-\sqrt{\pi})\ket{x}
\end{align}
\label{eq:S0}%
\end{subequations}
with $\Psi(x)=\bra{\Psi}x\rangle$ and $\mathbb{Z}$ denoting the integers. Note that the states $\ket{\psi_0}$ and $\ket{\psi_1}$ are not individually normalized but satisfy $\langle\psi_0|\psi_0\rangle + \langle\psi_1|\psi_1\rangle = \langle\Psi|\Psi\rangle = 1$. Tracing out the continuous mode leaves a qubit with density matrix
\begin{equation}
    \rho = \begin{pmatrix}\langle\psi_0|\psi_0\rangle & \langle\psi_1|\psi_0\rangle \\ \langle\psi_0|\psi_1\rangle & \langle\psi_1|\psi_1\rangle
    \end{pmatrix}.
\end{equation}
The fidelity to the target state $\hat{T}\ket{+}=(\ket{0} + e^{i\pi/4}\ket{1})\sqrt{2}$ is given by 
\begin{equation}
    F = \bra{+}\hat{T}^\dagger\rho\hat{T}\ket{+} = \frac{1}{2} + \Re\left(e^{i\pi/4}\bra{\psi_1}\psi_0\rangle\right). \label{eq:S2}
\end{equation}
We now calculate $\bra{\psi_1}\psi_0\rangle$ for $\ket{\Psi}=\hat{U}_T\ket{+_L}$. The normalized wave function of the approximate GKP state $\ket{+_L} = (\ket{0_L} + \ket{1_L})/\sqrt{2}$ for small $\Delta_x$ and $\Delta_p$ is given by \cite{gottesman2001encoding}
\begin{equation}
    \Psi_{\ket{+_L}}(x) = \frac{1}{\pi^{1/4}}\sqrt\frac{\Delta_p}{\Delta_x}\sum_{s}\exp\left(-\frac{(\sqrt{\pi}s)^2}{2}\Delta_p^2\right)\exp\left(-\frac{(x-\sqrt{\pi}s)^2}{2\Delta_x^2}\right).
\end{equation}
Multiplying $\hat{U}_T$ we get:
\begin{equation}
    \Psi_{\hat{U}\ket{+_L}}(x) = \frac{1}{\pi^{1/4}}\sqrt\frac{\Delta_p}{\Delta_x}\sum_{s}\exp\left[i\frac{\pi}{4}\left\{2\left(\frac{x}{\sqrt{\pi}}\right)^3 + \left(\frac{x}{\sqrt{\pi}}\right)^2 - 2\frac{x}{\sqrt{\pi}}\right\}-\frac{(\sqrt{\pi}s)^2}{2}\Delta_p^2-\frac{(x-\sqrt{\pi}s)^2}{2\Delta_x^2}\right]. 
\end{equation}
Using Eqs. \eqref{eq:S0} we can now calculate the overlap $\bra{\psi_1}\psi_0\rangle$, assuming negligible overlap between neighbouring peaks in the GKP wave function, which is valid when $\Delta_x^2\ll 1$:

\begin{align}
    \langle\psi_1|\psi_0\rangle=&\frac{1}{\sqrt{\pi}}\frac{\Delta_p}{\Delta_x}\sum_s\Bigg(\exp\left[-\frac{(2s\sqrt{\pi})^2}{2}\Delta_p^2\right]\exp\left[-\frac{((2s+1)\sqrt{\pi})^2}{2}\Delta_p^2\right]\nonumber\\
    &\times\int_{(2s-\frac{1}{2})\sqrt{\pi}}^{(2s+\frac{1}{2})\sqrt{\pi}}dx\,\exp\left[-i\frac{\pi}{4}\left(6\frac{x^2}{\pi}+8\frac{x}{\sqrt{\pi}}+1\right)\right]\exp\left[-\frac{(x-2s\sqrt{\pi})^2}{\Delta_x^2}\right]\Bigg). \label{eq:S1}
\end{align}
For $\Delta_x^2\ll 1$ we can expand the limits of the integrals to $\pm\infty$ and evaluate using the formula for Gaussian integrals,
\begin{equation}
    \int_{-\infty}^{\infty}dx\,e^{-ax^2+bx+c}=\sqrt{\frac{\pi}{a}}e^{c + \frac{b^2}{4a}}, \label{eq:GaussianInt}
\end{equation} 
where from Eq. \eqref{eq:S1} we identify
\begin{equation}
    a=\frac{1}{\Delta_x^2}+i\frac{3}{2},\qquad b=\frac{4s\sqrt{\pi}}{\Delta_x^2} - i2\sqrt{\pi}, \qquad \text{and}\qquad c=-\frac{4s^2\pi}{\Delta_x^2}-i\frac{\pi}{4}.
\end{equation}
Inserting and rewriting:

\begin{align}
    \langle\psi_1|\psi_0\rangle=&\frac{\Delta_p}{\Delta_x\sqrt{\frac{1}{\Delta_x^2} +i\frac{3}{2}}}\sum_s\Bigg(\exp\left[-4\pi\left(\frac{1}{\Delta_x^2} - \frac{1}{\Delta_x^2\left(1+\frac{9}{4}\Delta_x^4\right)} + \Delta_p^2\right)s^2 - 2\pi\left(\Delta_p^2 + \frac{3\Delta_x^2}{1 + \frac{9}{4}\Delta_x^4}\right)s - \pi\left(\frac{\Delta_x^2}{1 + \frac{9}{4}\Delta_x^4} + \frac{1}{2}\Delta_p^2\right)\right]\nonumber \\
    &\times\exp\left[-i\pi\left(\frac{1}{1+\frac{9}{4}\Delta_x^4}\left(6s^2+4s-\frac{3}{2}\Delta_x^4\right)+\frac{1}{4}\right)\right]\Bigg).
\end{align}
Now consider the term in the exponential of the last factor. For small $|s|$ we have $(6s^2+4s-(3/2)\Delta_x^4)/(1+(9/4)\Delta_x^4)\equiv 0 \text{ (mod 2)}$ and thus the last factor reduces to $\exp\left[-i\pi/4\right]$. If $\Delta_x^4\ll\Delta_p^2$ this holds for all non-vanishing terms in the sum. For $\Delta_x^2,\Delta_p^2\ll 1$ the terms in the sum then change slowly with $s$, and we can approximate the sum with an integral:

\begin{align}
    \langle\psi_1|\psi_0\rangle=\frac{\Delta_p e^{-i\frac{\pi}{4}}}{\Delta_x\sqrt{\frac{1}{\Delta_x^2} +i\frac{3}{2}}}\int ds\,\exp\left[-4\pi\left(\frac{1}{\Delta_x^2} - \frac{1}{\Delta_x^2\left(1+\frac{9}{4}\Delta_x^4\right)} + \Delta_p^2\right)s^2 - 2\pi\left(\Delta_p^2 + \frac{3\Delta_x^2}{1 + \frac{9}{4}\Delta_x^4}\right)s - \pi\left(\frac{\Delta_x^2}{1 + \frac{9}{4}\Delta_x^4} + \frac{1}{2}\Delta_p^2\right)\right].
\end{align}
The integral can again be evaluated using Eq. \eqref{eq:GaussianInt} with
\begin{equation}
    a = 4\pi\left(\frac{1}{\Delta_x^2} - \frac{1}{\Delta_x^2\left(1+\frac{9}{4}\Delta_x^4\right)} + \Delta_p^2\right), \qquad b = -2\pi\left(\Delta_p^2 + \frac{3\Delta_x^2}{1 + \frac{9}{4}\Delta_x^4}\right)\qquad \text{and} \qquad c = -\pi\left(\frac{\Delta_x^2}{1 + \frac{9}{4}\Delta_x^4} + \frac{1}{2}\Delta_p^2\right).
\end{equation}
Before inserting, we consider the limit of small $\Delta_x$ and $\Delta_p$ in which we get 
\begin{align}
    a &\rightarrow \pi(9\Delta_x^2 + 4\Delta_p^2),\\
    b &\rightarrow \pi(6\Delta_x^2 + 2\Delta_p^2),\\
    c &\rightarrow \pi(\Delta_x^2 + \frac{1}{2}\Delta_p^2),\\
    e^{\frac{b^2}{4a}}&\rightarrow 1,\\
    e^{c}&\rightarrow 1.
\end{align}
Evaluating the integral we thus get:
\begin{equation}
    \langle\psi_1|\psi_0\rangle = e^{-i\frac{\pi}{4}}\frac{\Delta_p}{\sqrt{9\Delta_x^2+4\Delta_p^2}\sqrt{1+i\frac{3}{2}\Delta_x^2}} \underset{\Delta_x\rightarrow 0}{\rightarrow} e^{-i\frac{\pi}{4}}\frac{\Delta_p}{\sqrt{9\Delta_x^2+4\Delta_p^2}}.
\end{equation}

Inserting in Eq. \eqref{eq:S2}:
\begin{equation}
    F = \frac{1}{2} + \frac{\Delta_p}{\sqrt{9\Delta_x^2+4\Delta_p^2}} = \frac{1}{2} + \frac{1}{2}\frac{1}{\sqrt{1+\left(\frac{3\Delta_x}{2\Delta_p}\right)^2}}. 
\end{equation}

\end{widetext}

\end{document}